\documentclass[%
 reprint,
 amsmath,amssymb,
 aps,
]{revtex4-2}

\usepackage{graphicx}
\usepackage{dcolumn}
\usepackage{bm}
\usepackage{hyperref}
\usepackage{xcolor}
\usepackage{soul}

\begin{document}

\preprint{APS/123-QED}

\title[RodsSiloRot]{Discharge of elongated grains in silos under rotational shear}

\author{Tivadar Pong\'o,\textit{$^{1,2}$}
Tam\'as B\"orzs\"onyi\textit{$^{2}$}
and 
Ra\'ul Cruz Hidalgo,\textit{$^{1}$}
} 

\address{
$^1$F\'isica y Matem\'atica Aplicada, Facultad de Ciencias, Universidad de Navarra, Pamplona, Spain \\
$^2$Institute for Solid State Physics and Optics, Wigner Research Centre for Physics, P.O. Box 49, H-1525 Budapest,
}

\date{\today}

\begin{abstract}
The discharge of elongated particles from a  silo with rotating bottom is investigated numerically. The introduction of a slight transverse shear reduces the flow rate $Q$ by up to 70\% compared to stationary bottom, but the flow rate shows a modest increase by further increasing the external shear.
Focusing on the dependency of flow rate $Q$ on orifice diameter $D$, the spheres and rods show two distinct trends. 
For rods, in the small aperture limit $Q$ seems to follow an exponential trend, deviating from the classical power-law dependence. These macroscopic observations are in good agreement with our earlier experimental findings [\href{https://doi.org/10.1103/PhysRevE.103.062905}{\color{blue}{Phys.~Rev.~E~\textbf{103}, 062905 (2021)}}].
With the help of the coarse-graining methodology we obtain the spatial distribution of the macroscopic density, velocity, kinetic pressure, and orientation fields. This allows us detecting a transition from funnel to mass flow pattern, caused by the external shear.
Additionally, averaging these fields in the region of the orifice reveals that the strong initial decrease in $Q$ is mostly attributed to changes in the flow velocity, while the weakly increasing trend at higher rotation rates is related to increasing packing fraction.
Similar analysis of the grain orientation at the orifice suggests a correlation of the flow rate magnitude with the vertical orientation and the packing fraction at the orifice with the order of the grains. Lastly, the vertical profile of mean acceleration at the center of the silo denotes that the region where the acceleration is not negligible shrinks significantly due to the strong perturbation induced by the moving wall.
\end{abstract}

\maketitle

\section{\label{sec:introduction}Introduction}

Granular materials are everywhere in nature and they are often used in industrial processes.
Since long time, humans have employed containers like
silos and bins to store them,
so it is technologically important to understand their mechanical response under these specific boundary conditions.
Thus, notable research efforts have been undertaken in this direction, where the  ultimate aim is to derive the macroscopic response of a granular sample from the contact interactions of the whole particle ensemble \cite{Nedderman1992, Andreotti2013}. 

Developing technological solutions, several types of  silo flow patterns have been detected. For instance, a  {\it funnel flow} pattern is characterized by the initial particle flow in the central region of the silo. 
Consequently, funnel flow silos often have stagnant grains near the walls, leading to undesired in-service issues.  
In contrast, {\it mass flow} pattern provides a uniform outflow
without a central flow channel, 
and the material flows down as a continuum column.
Achieving mass flow condition is ideal, in particular, for mixtures of particles that are susceptible to segregation. 

The dependence of the particle flow rate $Q$ on the orifice diameter $D$ also has a significant technological interest. Due to its simplicity, the most used expression is the well-known Beverloo's correlation: $Q \propto (D - kd)^{5/2}$ \cite{beverloo}.  
In the formulation $d$ is the particle diameter and the parameter $k$ enables a good fit of the experimental data for small orifices. In the large-orifice limit ($D \gg kd$), 
however, a simple power-law $Q \propto D^{5/2}$ is obtained. 
Recently, Janda et al.~presented a different approach to predict the particle flow rate \cite{Jandaprl2012}.
Examining the kinetic spatial profiles of density and velocity at the orifice of a two-dimensional (2D) silo, they obtained self-similar functions, when using the orifice size $D$ as a characteristic length.
This analysis led them to the formulation of the expression $Q \propto (1 - \alpha_1 e^{-D/\alpha_2}) D^{5/2}$ in which the term in the parentheses accounts for the scaling of the packing fraction and thus the density.
They found the fitting parameters $\alpha_1$, $\alpha_2$ to be around $1/2$, and 6 particle diameters, respectively.
In all the described approaches, the value of the exponent $5/2$ can be justified by arguing that once a particle reaches a distance $D$ to the orifice, it starts accelerating. The region of accelerating flow is historically known as the {\it free fall arch region}.

To control the outflow in silos, several approaches have been used. Typically, the silos and hoppers have been perturbed, for instance, using electric fields to control the outflow of metallic beads \cite{Chen2001}, magnetic fields to introduce vibrations in the orifice region \cite{Dave2000}, or inducing a repulsive force between the grains \cite{Hernandez-Enriquez2017, Thorens2021}. The impact of external vibration on the macroscopic flow rate of a silo is far from being understood. More than thirty years ago, Hunt et al. experimentally observed a flow rate enhancement when increasing the intensity of a horizontal vibration \cite{Hunt1999}. Vertical vibrations, however,  produce a more complex response, showing a flow rate decrease when rising the dimensionless acceleration amplitude $\Gamma$.  On the contrary, when using a higher oscillation frequency, $f$, a slight increase of $Q$ against the same parameter was encountered \cite{Wassgren2002}. Pascot et al.  have recently obtained experimentally and numerically the existence of two different regimes when varying the oscillation amplitude $A$, and fixing the frequency \cite{Pascot2020}. In particular, when analyzing small orifices, it is accepted that introducing vibrations alters the arches' stability \cite{jandaEPL2009, Mankoc2009},  and the distributions of the unclogging times \cite{Merrigan2018, Guerrero2019}. 

Imposing an external shear is also a promising alternative to avoid the formation of stable arches in silos. In the past, the discharge of a cylindrical silo with rotating bottom was explored \cite{Corwin2008}. However,  the authors only focused their attention on the dynamics of the system's surface \cite{Corwin2008}. Later on, Hilton and Cleary \cite{Hilton2010} examined the impact of the external shear on the flow rate $Q$, finding that it is unaffected when a low shear rate is applied. However, after reaching a critical value, $Q$ increases monotonically with the rotational frequency.

Very recently, the discharge of wooden rods from a cylindrical silo perturbed by a rotating bottom wall was investigated experimentally \cite{to2021discharge}. It was mainly found that, for small orifices,  the flow rate deviates from the classical power-law correlation $Q\sim D^{5/2}$, and an exponential dependence $Q \sim e^{\kappa D}$ is detected.
More interestingly, in the continuous flow regime, the introduction of transversal shear induced by the bottom wall's movement decreases the flow rate significantly \cite{to2021discharge} compared to spheres with similar effective dimensions \cite{to2019granular}. Further increasing the rotation rate results in an increase in the flow rate, which is more significant for smaller orifice sizes, where the flow is intermittent. Since elongated particles in a shear flow develop orientational ordering \cite{borzsonyi2012orientational, borzsonyi2012shear}, it is expected that changes in particle orientations due to the external shear will alter the discharge rate.
The complex behavior observed in the experiments was not fully explained since the experimental approach has the limitation of not accessing the velocity field, the density distribution and the particle orientations inside the silo.

In this work, we introduce a numerical approach, which replicates the experimental scenario examined in Ref.~\cite{to2021discharge}, reproducing their main macroscopic observations. Our aim is to extract relevant features of the process, which were not accessible experimentally. Thus, our approach shed light on understanding, how silo discharge of nonspherical grains is influenced by an external transversal shear. 

\section{Numerical Model}
\label{numeric}
\begin{figure}
    \centering
    \includegraphics[width=1.0\columnwidth]{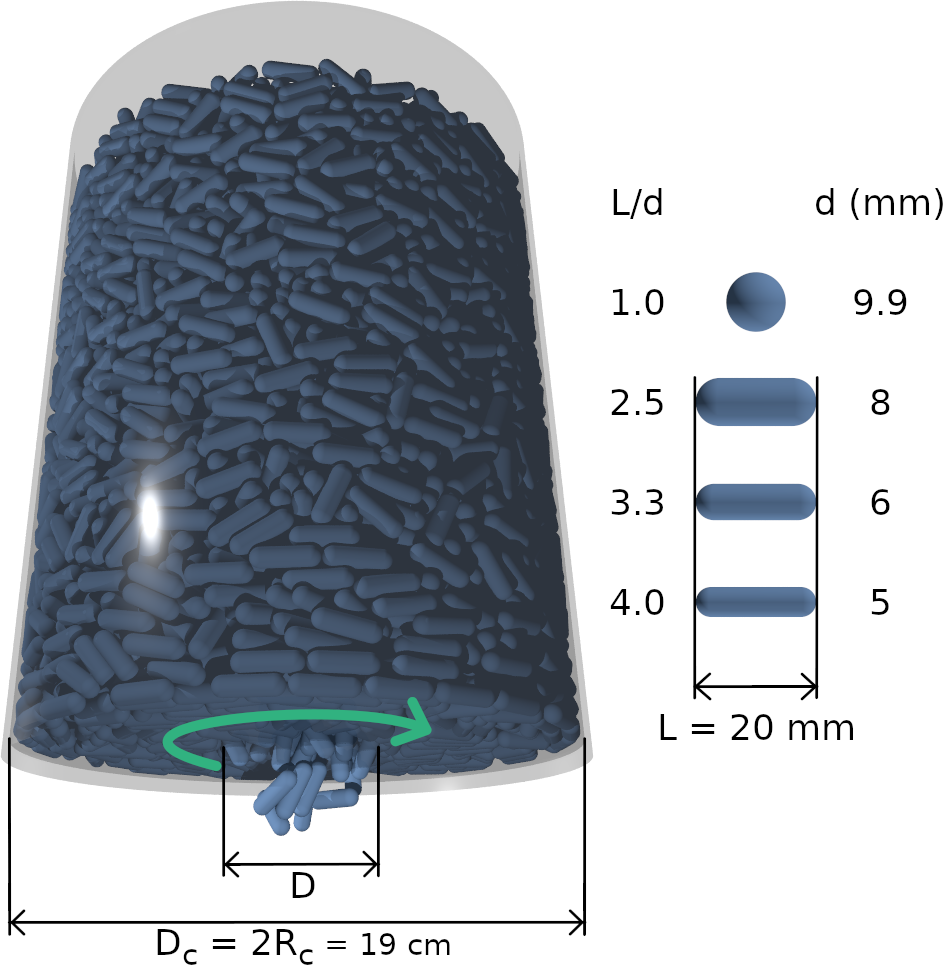}
    \caption{Visualization of the rotating bottom silo setup during the discharge of elongated particles of $L/d = 3.3$ (left).
    The system dimension replicates the experimental setup examined in Ref.~\cite{to2021discharge}.
    On the right, we display the spherocylinders used within this work, together with their corresponding dimensions. }
    \label{fig:setup}
\end{figure}

The numerical simulation consists in, using a Discrete Element Method (DEM) implementation \cite{sara2016}, modeling the mechanical behavior of elongated particles. The modeled system is a cylindrical flat silo with a diameter of $D_c = 2 R_c= 19$ cm, and bottom wall with a circular orifice in the center, with diameter $D$. As a novelty, an external transversal shear is imposed by the rotation of the silo bottom wall (see Fig.~\ref{fig:setup}).   

We examine the silo discharge, varying systematically the particle elongation. For simplicity, we employ spherocylinders to model the experimental wooden rods (see Ref.~\cite{to2021discharge}).
The model defines pairwise forces between contacting particles. The particle-particle interaction is computed using an algorithm for interacting spheropolyhedra \cite{marroquin08a,marroquin09a}; in particular, an implementation on GPU architecture \cite{sara2016}.  A spherocylinder is defined by the location of its two vertices and its sphero-radius $r=d/2$.
The spherocylinder's surface is depicted by the set of points at a distance $r$ from the segment which connects its two vertices.
The contact detection between two spherocylinders involves finding the closest point between the two edges \cite{vega1994fast}. The overlap distance $\delta_n$ between two spherocylinders equals the overlap distance of two spheres of radius $r$, located at each edge.
Then, the contact force is determined as the sum of the normal and tangential components $\boldsymbol{F}_{ij} = F_{n} \hat{\boldsymbol{n}} + F_{t} \hat{\boldsymbol{t}}$.
The normal unit vector $\hat{\boldsymbol{n}}$ points from the center of one sphere to another, while the tangential $\hat{\boldsymbol{t}}$ is parallel to the tangential velocity.
The normal force is calculated by a simple linear spring-dashpot model $F_{n} = -k_n \delta_n - \gamma_n v_n^{\text{rel}}$ where $k_n$ is a spring coefficient, $\gamma_n$ a damping coefficient, $\delta_n$ is the overlap between two particles, and $v_n^{\text{rel}}$ is the relative velocity in the normal direction.
The tangential force represents the friction between two particles, it is also modeled with a spring-dashpot but taking into account the constraint of Coulomb friction: $F_{t} = \min{ \{ -k_t \delta_t - \gamma_t v_t^{\text{rel}} , \mu F_{n}\} }$.
Here the first term represents a spring with coefficient $k_t$ and tangential deformation $\delta_t = |\boldsymbol{\delta}_t|$  which is integrated from the following $d \boldsymbol{\delta}_t/ dt = \boldsymbol{v}_t^{\text{rel}}$ where $\boldsymbol{v}_t^{\text{rel}}$ is the relative tangential velocity. The vector $\boldsymbol{\delta}_t$  is kept parallel to the contact plane by truncation \cite{weinhart2012closure}.
The second term is proportional to the tangential velocity with a coefficient $\gamma_t$, while the particle friction $\mu$ sets the constraint for the force.

The DEM algorithm integrates the equations of motion for both translational and rotational degrees of freedom, accounting for gravity $\boldsymbol{g}$ and the force $\boldsymbol{F}_{ij}$ acting between contacting particles.
A velocity Verlet method \cite{hairer2003geometric} is  used to integrate the translational, and a modified leapfrog \cite{wang2006implementation} resolves the rotational ones.

As initial conditions, packing of monodisperse rods were created by letting a dilute granular gas settle inside the closed silo by gravity $g = 9.8$ m/s. The system is composed of $N = 10 000$ particles, and only in the case of $L/d = 4.0$ $N$ is set to $22500$ to reach a bit higher initial packing (note that in this case the particle's volume is smaller). The discharge process starts by opening the orifice and simultaneously setting the rotation frequency of the bottom wall to one of the following values: $f = 0.0, 0.16, 0.32, 0.64, 0.96,$ and $1.28$ Hz. Simulation time for the discharge was in the range of 1--7 hours depending on the flow rate, thus overall discharge time, particle shape and the type of GPU used (NVIDIA GeForce RTX 2070 and 3080).

The simulation mimics the behavior of wooden particles, thus, 
the particle density $\rho_p$ is set to $620 \text{kg}/\text{m}^3$, a spring stiffness $k_n = 2\cdot 10^5 ~ m_p g /d$ is used \cite{silbert2001granular}. Other model parameters were $e_n = 0.9, k_t = 2/7 k_n, \gamma_t = \gamma_n$ and $\mu = 0.5$. The normal damping constant is dependent on the coefficient of restitution $e_n$ in the following way: $\gamma_n = \sqrt{\frac{2k_n m_p}{\left(\pi/\ln{(e_n)}\right)^2} + 1}$ \cite{schwager2007coefficient}. The time step is $\Delta t = 10^{-6}$ s, which is smaller than 2\% of the contact time between two colliding particles \cite{huang2014time}.

The DEM algorithm provides the evolution of the trajectories of all the particles, and their contact network, with the desired time resolution.
Post-processing this data employing a coarse-graining methodology enables a well-defined continuous description of the granular flows, via the packing fraction 
\begin{equation}
    \phi(\boldsymbol{r},t) = \frac{1}{\rho_p} \sum_i^N m_i \varphi(\boldsymbol{r} - \boldsymbol{r}_i(t)),
\end{equation}
linear momentum $\boldsymbol{P}(\boldsymbol{r},t)$, and velocity 
\begin{equation}
\boldsymbol{V}(\boldsymbol{r},t) = \sum_i^N \boldsymbol{v}_i \varphi(\boldsymbol{r} - \boldsymbol{r}_i(t))
\end{equation}
fields \cite{BABIC1997,Goldhirsch2010}. 
Here $m_i$, $\boldsymbol{v}_i$ are the mass and velocity of particle $i$, and $\varphi(\boldsymbol{r})$ is non-negative integrable function that serves to coarse-grain our particles.
In particular, we use a truncated Gaussian function $\varphi(\boldsymbol{r}) = A_{\omega}^{-1} \exp[\boldsymbol{r}^2/2\omega^2]$, with $A_{\omega}^{-1}$ chosen so that the integral of $\varphi(\boldsymbol{r})$ over all the space results in 1, the cutoff distance is $r_c = 4\omega$ and $\omega = d^*/2$, where $d^*$ is the sphere equivalent diameter $d^*  = \large( \frac{3V_p}{4\pi} \large)^{\frac{1}{3}}$, with $V_p$ corresponding to the volume of the particle.
Additionally, the contact $\sigma^c(\boldsymbol{r},t)$ and kinetic $\sigma^k(\boldsymbol{r},t)$ stress tensor fields are calculated in the following way:
\begin{equation}
    \sigma_{\alpha \beta}^c (\boldsymbol{r},t) = - \frac{1}{2} \sum_{i,j}^{N} 
    f_{ij \alpha}r_{ij \beta} \int_0^1 \varphi(\boldsymbol{r} - \boldsymbol{r}_i + s \boldsymbol{r}_{ij}) \,\mathrm{d}s,
\end{equation}
\begin{equation}
    \sigma_{\alpha \beta}^k (\boldsymbol{r},t) = - \sum_{i}^N m_i v'_{i\alpha} v'_{i\beta} \varphi(\boldsymbol{r} - \boldsymbol{r}_i).
\end{equation}
While the former is computed by a line integral over $\boldsymbol{r}_{ij} = \boldsymbol{r}_{i} - \boldsymbol{r}_{j}$ considering the force acting between particles $i$ and $j$, the kinetic stress represents a granular temperature due to the multiplication of the velocity fluctuation terms $\boldsymbol{v}'_i (\boldsymbol{r}, t) = \boldsymbol{v}_i (t) - \boldsymbol{V}(\boldsymbol{r}, t) $.
In the case of elongated particles, the components of the mean orientational tensor $O$ also provide very useful information during the analysis.
We have used the following formula for its coarse-graining:
\begin{equation}
    O_{\alpha \beta} (\boldsymbol{r},t) = \sum_i^N l_{i\alpha} l_{i\beta} \varphi(\boldsymbol{r} - \boldsymbol{r}_i),
\end{equation}
where $\boldsymbol{l}_i$ is the unit vector representing the particle's direction. By definition, the diagonal elements of the orientation tensor are non-negative and fall between $0$ and $1$, representing the degree of alignment in the specific direction.
In this way, the quantity expressing the nematic order $S$ of the ensemble of particles is the largest eigenvalue of the $O$ tensor. With our convention $S$ takes values from $1/3$ (completely disordered) up to $1$ (fully ordered). In this work, we discuss the orientation of the particles in the cylindrical coordinate system due to the specific symmetry of the silo. Furthermore, for all the coarse-grained quantities we applied an averaging in the azimuthal direction in the following way: $ X(r, z, t) = \frac{1}{2\pi} \int_0^{2\pi} X(\boldsymbol{r}, t) \, \mathrm{d}\theta $.
More details about the data post-processing  can be found in  Ref.~\cite{BABIC1997,Goldhirsch2010,zhang2010coarse,patrick05a,weinhart2012closure}.

In our analysis, we are interested in steady-state conditions, namely, the part of the process in which the flow is stationary. 
In the case of the static bottom, the discharge slows down near the end of the process due to the lack of particles outside the stagnant zone. While in the case of the quickly rotating bottom, when the height of the material in the silo goes below about $R_c$, the whole ensemble starts spinning faster. To exclude these initial and final effects, we focus on a time interval, where the flow is stationary in all cases.
To have a better statistics, we ran simulations for each set of parameters, starting from three slightly different initial configurations, and apply time and ensemble averaging. Different seeds have been used for the random generator in order to construct different dilute granular gases at the beginning, which yielded distinct initial packings.
In the case of small orifices with rotating bottom, we observed intermittent flow. To handle this, during the calculation of the flow rate we excluded the intervals larger than $1$ s where there was no flow.

\section{Results}
\subsection{Effect of rotational shear on the flow rate}

Using a numerical approach, we aim to elucidate how the external transversal shear introduced by the bottom wall rotation impacts the discharge process.
As mentioned earlier, a recent experimental work examined the flow of particles with aspect ratio $L/d = 3.3$ and $L/d=2.5$. Here, we extend these findings by systematically varying $L/d$ over a wider range. In line with the experiments, the length of the spherocylinder  (measured from end-to-end) is the same $L = 20$ mm, in all the cases. 
However, the aspect ratio $L/d$ is varied by changing the spherodiameter $d$. In particular, we investigate the behavior of grains with the following aspect ratios: $L/d = 1.0, 2.5, 3.3,$ and $4.0$, whereby $L/d = 1.0$ refers to spherical particles. 
 The diameter of the spherical particle is chosen so that its volume would be equal to the volume of a rod with $L/d = 3.3$: $d = d^*_{L/d=3.3} = 9.9$ mm, which is also the definition for their equivalent diameter.

\begin{figure}
    \centering
    \includegraphics[width=1.0\columnwidth]{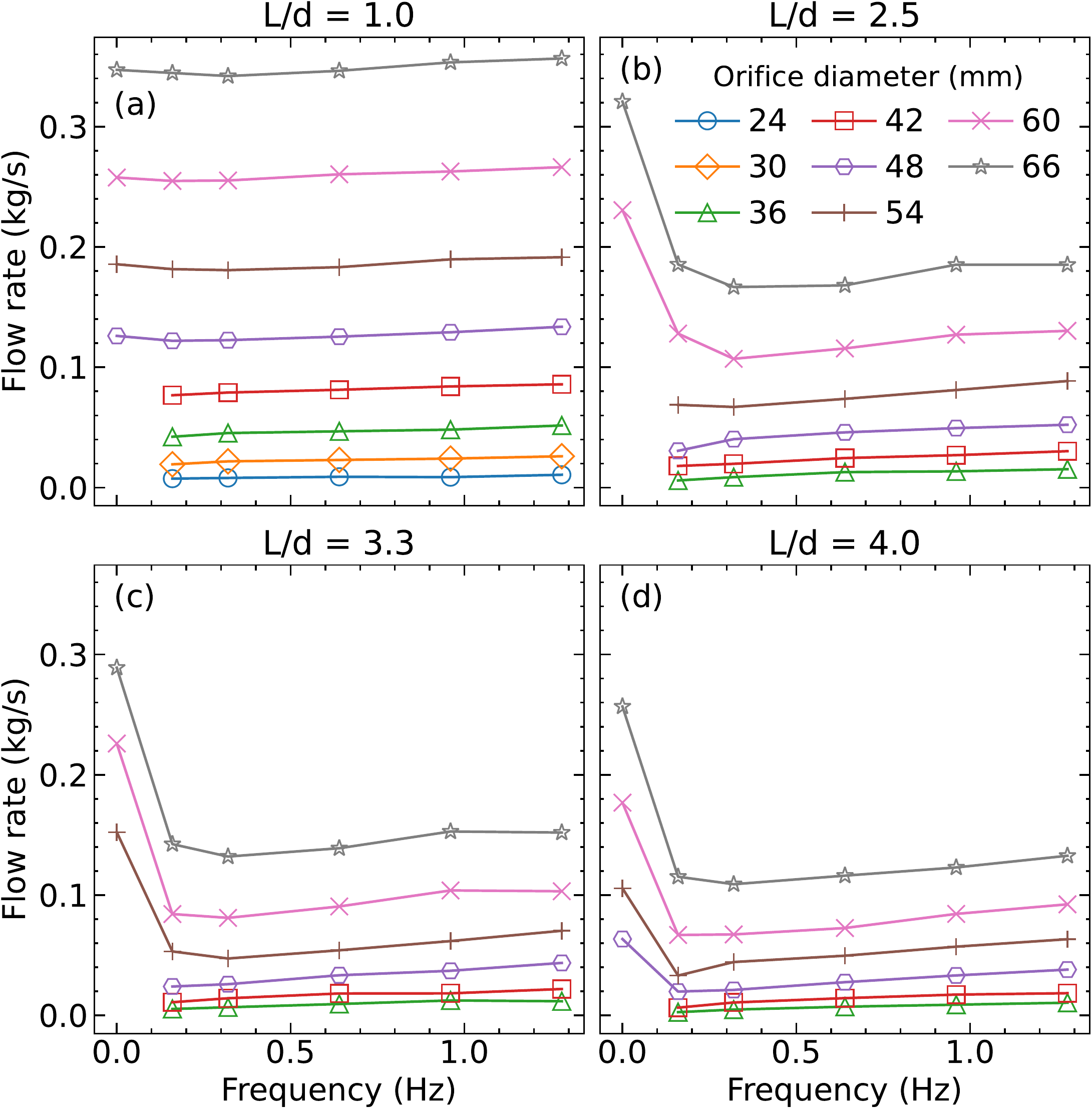}
    \caption{Discharge rate as a function of the base rotation frequency. The different panels show data for (a) spheres, and (b)-(d) spherocylinders with aspect ratios of $L/d = 2.5, 3.3, 4.0$ respectively. Each curve corresponds to a specific orifice diameter displayed in the legend.}
    \label{fig:flowrate_vs_freq}
\end{figure}

In Fig.~\ref{fig:flowrate_vs_freq} we present the flow rate $Q$ as a function of the bottom rotation frequency, obtained for several aspect ratios and orifice sizes.
For comparison, the first panel includes the results corresponding to spheres with diameter $d=9.9$ mm. As expected, we find that for small orifice diameters ($D/d^* \lesssim 5$), a stationary flow does not develop, in any of the cases. However, the imposition of an external transversal shear induces the particle flow. For the case of the spheres, we obtain a slight increase with increasing rotational frequency below $D = 48$ mm or even a weak nonmonotonic behavior ($\approx 5\%$ initial drop then $\approx 5\%$ increase) for the larger orifices which we consider.
A more in-depth analysis of the behavior of the spheres in this system can be found in our earlier works \cite{to2019granular,hernandez2020particle}.  
In contrast, elongated particles show a completely different behavior, regardless the particle aspect ratio, either the orifice size, applying external shear significantly reduces the particle flow rate in the range of examined frequencies. It is worth mentioning that at faster rotations the increase of the rotation frequency $Q$ leads to a slight increase of the flow rate.   
Remarkably, these results nicely agree with the experimental analysis executed using wooden rods \cite{to2021discharge}.

\begin{figure}
    \centering
    \includegraphics[width=\columnwidth]{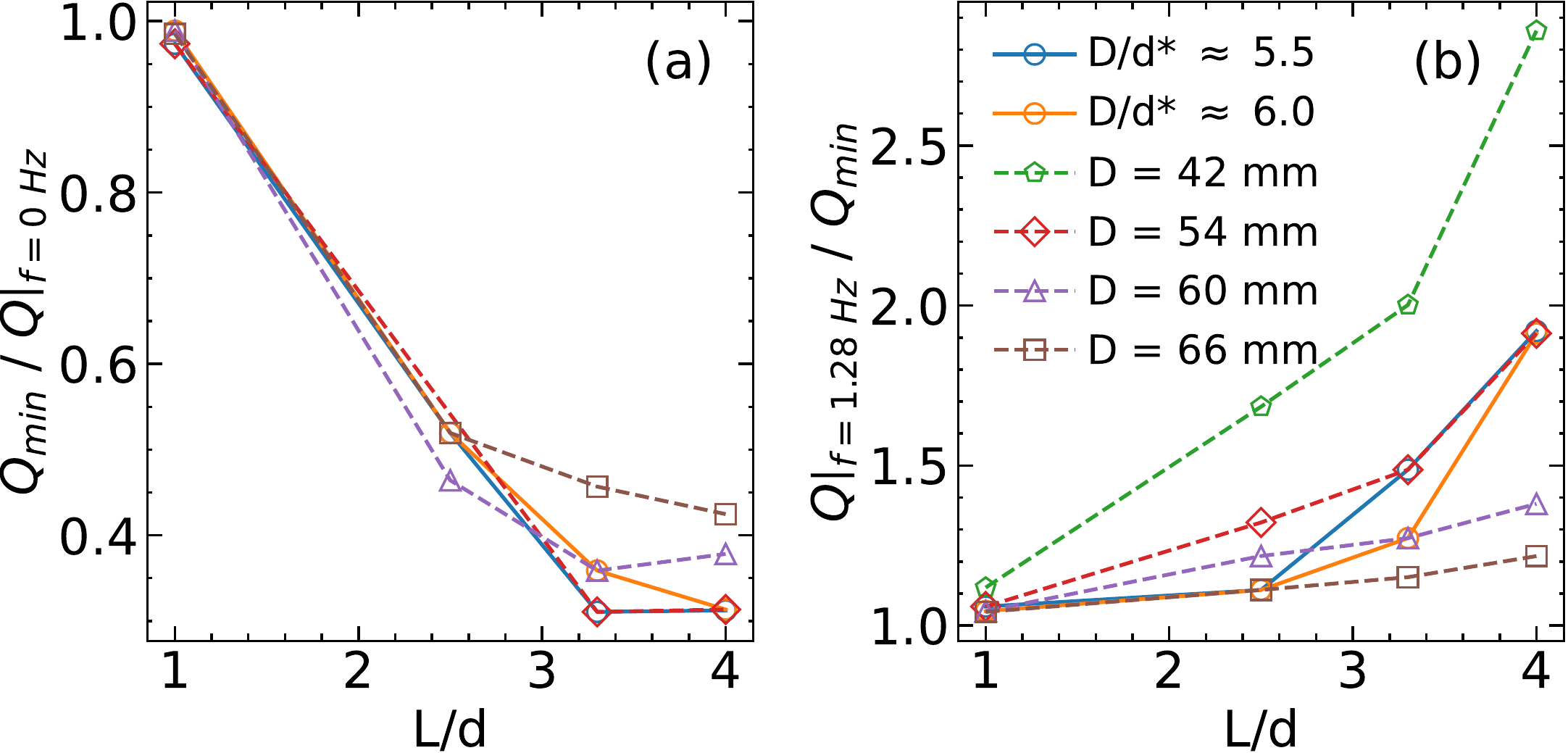}
    \caption{
    (a) Relative initial flow rate drop and (b) relative flow rate increase as a function of the particle aspect ratio. Here $Q_{\text{min}}$ denotes the minimum of the flow rate in the frequency variable, $\left. Q \right|_{f=0\ \text{Hz}}$ and $\left. Q \right|_{f=1.28\ \text{Hz}}$ are the flow rates measured with the stationary bottom and with the fastest rotation rate, respectively. Lines correspond to data with fixed relative orifice diameter $D/d^*$, dashed lines display data for fixed absolute diameter sizes.
    }
    \label{fig:flowrate_drop}
\end{figure}

Next, we extend the analysis, exploring the impact of the particle effective diameter $d^*$.
The data presented in Fig.~\ref{fig:flowrate_vs_freq} suggest that $Q$ generally reaches a minimum value in the range of rotational frequency of $0.16-0.32$ Hz. To quantify the drop, in Fig.~\ref{fig:flowrate_drop}(a), we plot the ratio of the minimum flow rate $Q_{\text{min}}$ and the one corresponding to the stationary bottom $\left. Q \right|_{f=0\ \text{Hz}}$. A general trend becomes apparent: the increasing aspect ratio of the particles causes a larger decrease in the flow rate.
After the initial decrease in $Q$, an increase is observed for fast rotation speeds. In Fig.~\ref{fig:flowrate_drop}(b), we quantify this increase as the ratio of the flow rate at $f=1.28$ Hz and at the minimum. This clearly shows that the relative increase gets larger as the grains are more elongated or as the orifice size is decreased.
Since our particles have different effective diameters, we include two groups of curves in the figure: one with fixed absolute orifice sizes (dashed curves) and one with fixed orifice sizes relative to the effective diameter of the particles (lines). In both cases, our findings are very conclusive; the particle aspect ratio has a significant impact on the system flow rate, which could drop by even $70\%$ for very elongated grains. 

\begin{figure}
    \centering
    \includegraphics[width=0.9\columnwidth]{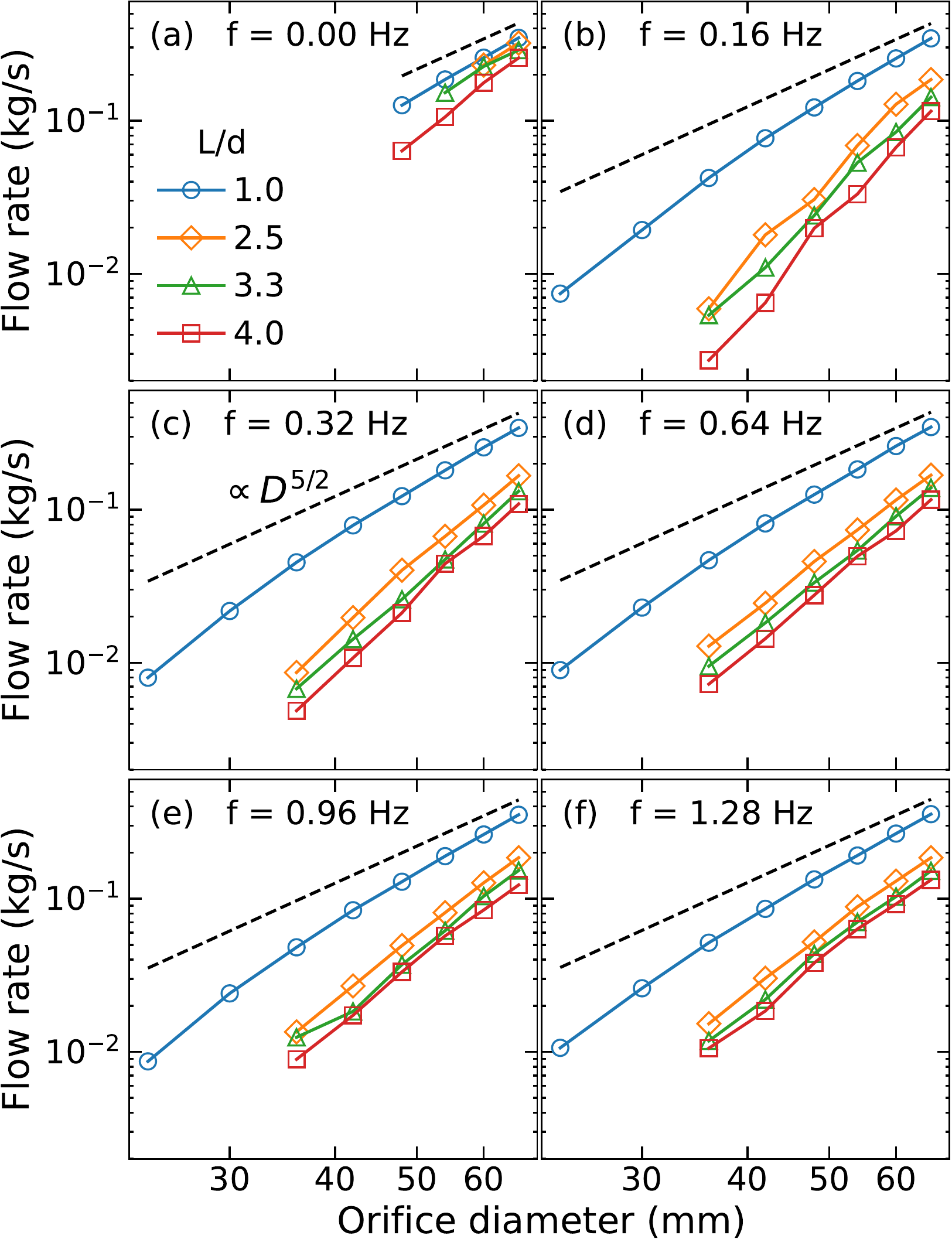}
    \caption{Log-log plot of the discharge rate as a function of orifice size.
    The different panels (a)-(f) correspond to different bottom rotation frequencies $f = 0, 0.16, 0.32, 0.64, 0.96, 1.28$ Hz respectively, as indicated inside them, while each curve presents data for a specific particle aspect ratio shown in the legend. The back dashed line represents a power-law function with an exponent of $5/2$.}
    \label{fig:flowrate_vs_D_loglog}
\end{figure}

\begin{figure}
    \centering
    \includegraphics[width=0.9\columnwidth]{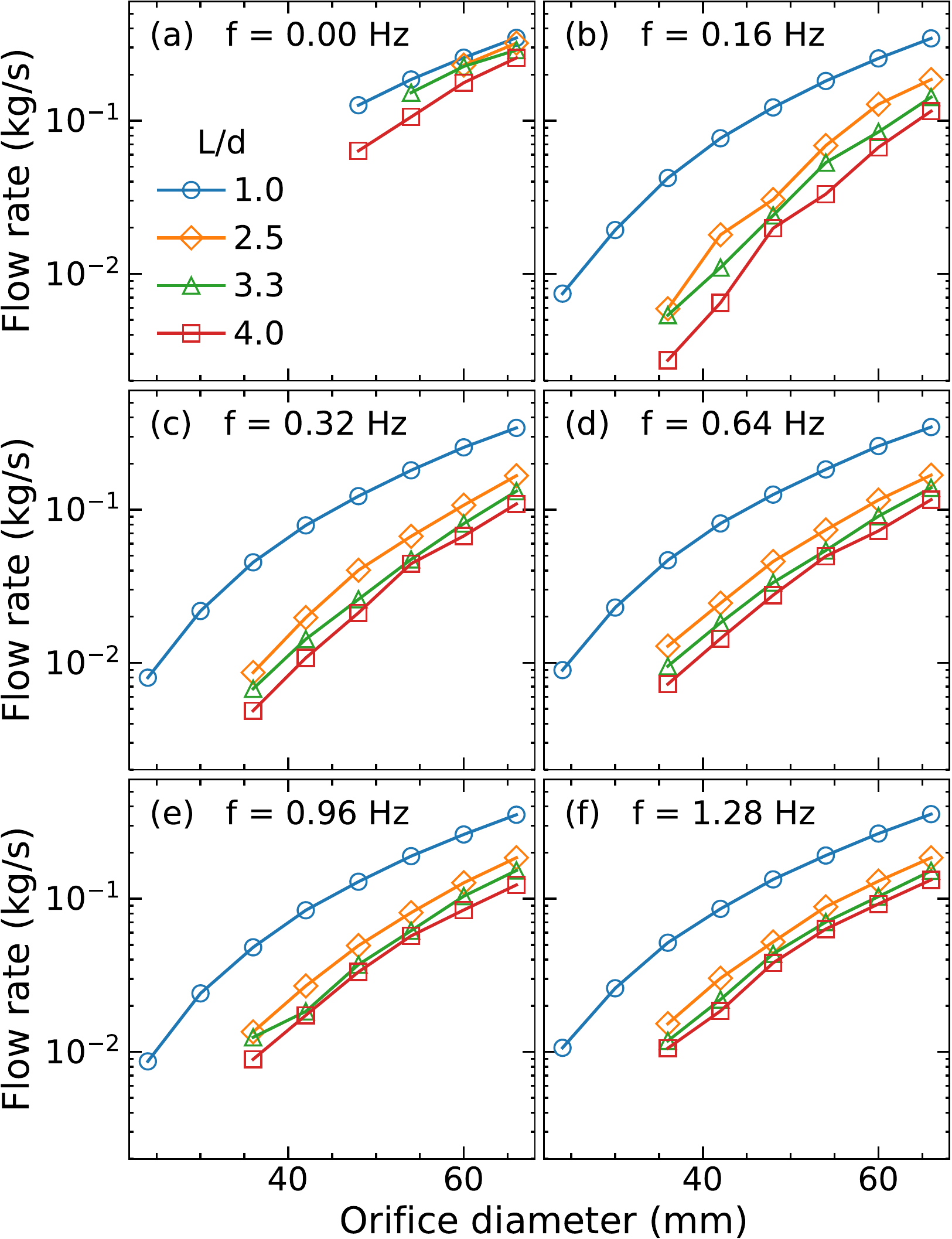}
    \caption{Semilog plot of the discharge rate as a function of orifice size.
    The different panels (a)-(f) correspond to different bottom rotation frequencies $f = 0, 0.16, 0.32, 0.64, 0.96, 1.28$ Hz respectively, as indicated inside them, while each curve presents data for a specific particle aspect ratio shown in the legend.}
    \label{fig:flowrate_vs_D_log}
\end{figure}

In the following, we examine the dependence of the flow rate $Q$ on the orifice size $D$. In Fig.~\ref{fig:flowrate_vs_D_loglog}, we present data  $Q$ vs $D$, in a log-log plot; additionally,  a dashed line representing the classic power-law function with an exponent of $5/2$ is included.  
One visible tendency is that the more elongated the particles are, the lower the flow rate for a fixed orifice is. 
For the case of spheres, we observe a concave line in the plots, which is in agreement with earlier findings \cite{to2019granular, hernandez2020particle}. This behavior can be explained by either the Beverloo law $Q \propto (D-kd)^{5/2}$ or $Q \propto (1-\alpha_1 e^{-D/\alpha_2}) D^{5/2}$ type of dependence. 
In the cases of rods, in the limit of small orifices, 
we obtain power-law relation but with a larger exponent, 
although in the limit of large orifices, the power-law correlation $Q \sim D^{\frac{5}{2}}$ is also reproduced.   
Complementarily, in Fig.~\ref{fig:flowrate_vs_D_log} we present the flow rate data in semilog plots. It highlights that the flow rate is also compatible with an
exponential trend $Q\sim e^{\kappa D}$, when decreasing the orifice diameter. 
In general, the value of $\kappa_{\rm sim} = 0.084$ mm$^{-1}$, fits with a good accuracy our observations. 
Remarkably, a similar trend was obtained experimentally in Ref.~\cite{to2021discharge}, reporting $\kappa_{\rm exp} = 0.13$ mm$^{-1}$. However, it is worth mentioning that the range of examined orifice sizes is narrower than in the experiments \cite{to2021discharge}, due to higher probability of clogging obtained numerically. 


\subsection{Continuum analysis of the particle flow}

\begin{figure}
    \centering
    \includegraphics[width=\columnwidth]{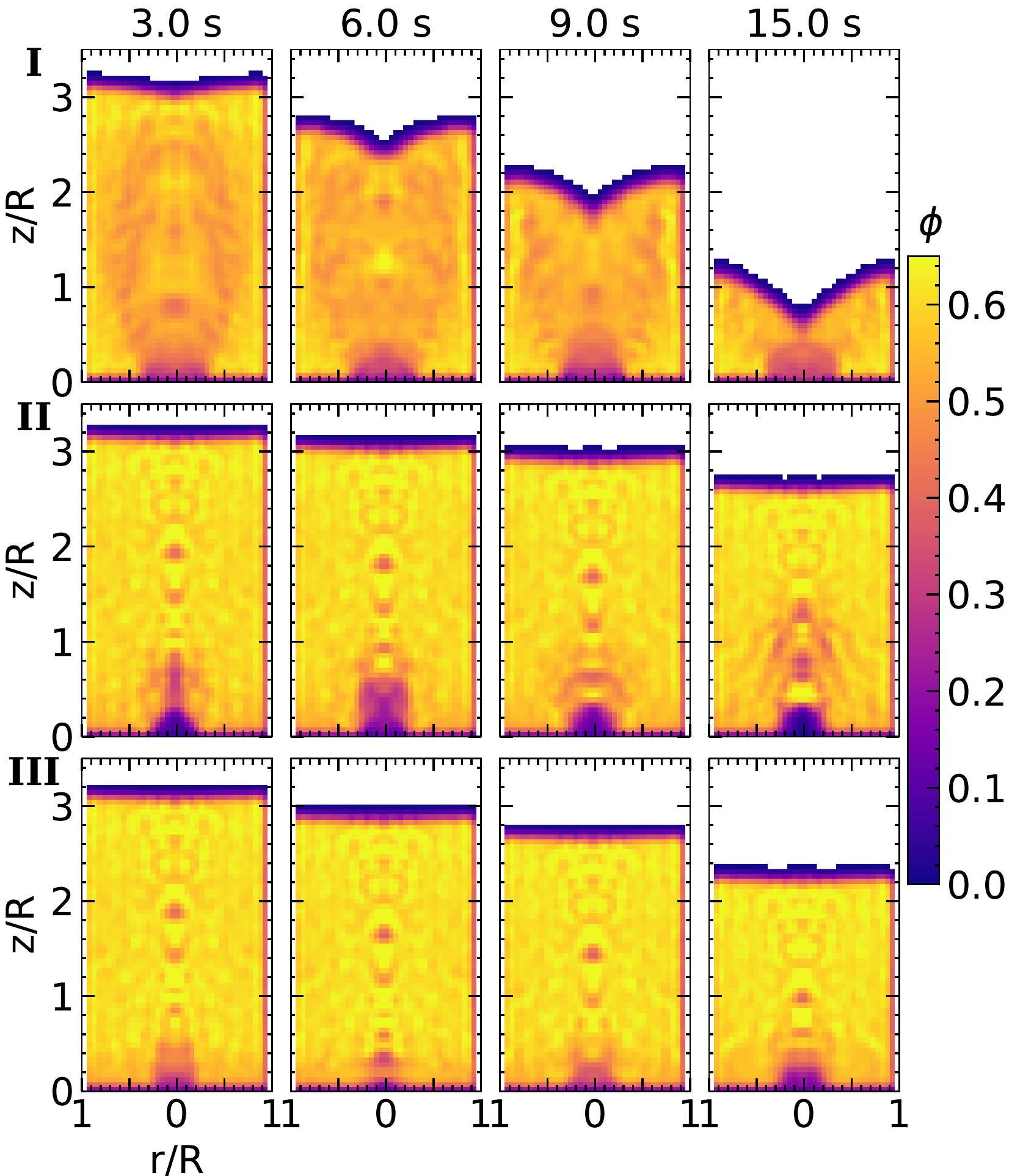}
    \caption{Color map representation of the spatial distribution of the packing fraction for elongated particles of $L/d = 3.3$, next to an orifice with size $D = 54 \,\text{mm}$. The rows correspond to $f$ = 0.0, 0.32, and 1.28 Hz in order. The different columns show the packing fraction in different instants of time as indicated above.}
    \label{fig:cg_phi}
\end{figure}

To gain insight into how the macroscopic properties of the flow are related to the flow patterns inside the silo, we employ a coarse-graining methodology, and derive the most relevant macroscopic fields from the data provided by the DEM simulations (see  Sec.~\ref{numeric}).
In Fig.~\ref{fig:cg_phi}, the colormaps represent the spatial profiles of the packing fraction $\phi(r,z)$. The data presented in rows I-III correspond to rotation frequencies $f = 0.0, 0.32,$ and $1.28$ Hz (up-down), respectively.  The graphs included in each column illustrate the system at different stages of the process, the indicated time is measured from the beginning of the discharge. Although the computation is done in cylindrical coordinates, the fields represent spatial averages in the azimuthal direction $\phi(r,z)$, within a time window of 1 s, $\pm 0.5$ s around the indicated time instant. Note that the fluctuations are larger in the center of the silo due to the cylindrical averaging having less data points there. Furthermore, the orifice region also shows larger noise; consequently, it should not be taken as representative of the whole discharge process.

As mentioned earlier, in Fig.~\ref{fig:cg_phi} the silo with a fixed bottom wall is again used as a baseline, which denotes the presence of {\it funnel type} of flow pattern. It is worth mentioning that this pattern is more noticeable when examining more elongated particles (see the Supplemental Material \cite{supplementarymat}).
This is characterized by a low-density region at the center of the silo and by the presence of a stagnant zone close to the lateral walls. Moreover, 
a surface depression is formed in the central region.  Interestingly, two significant changes occur when the bottom wall rotation is applied to the system: the top surface of the column becomes flat, and the funnel-shaped low packing fraction region disappears. A less visible change happens near the bottom wall, where the stagnant zone disappears, and the external shear causes a slight dilation (see below about $z/R < 0.4$). We also note that the low packing fraction region in the center near the orifice at $f = 1.28$ Hz shrinks compared to the $f=0.32$ Hz case. The spatial distributions of other particle elongations show qualitatively the same response to the rotational shear \cite{supplementarymat}.
Similar to previous studies of flow of elongated particles \cite{borzsonyi2012orientational,Tamas2017}, 
we also obtained a significant increase of the orientational ordering and the particle realignment in the sheared regions. 
These feactures were quantified using the orientational field $O_{zz}(r,z)$ and nematic order $S(r,z)$ fields
(see Supplemental Material \cite{supplementarymat}).

\begin{figure}
    \centering
    \includegraphics[width=\columnwidth]{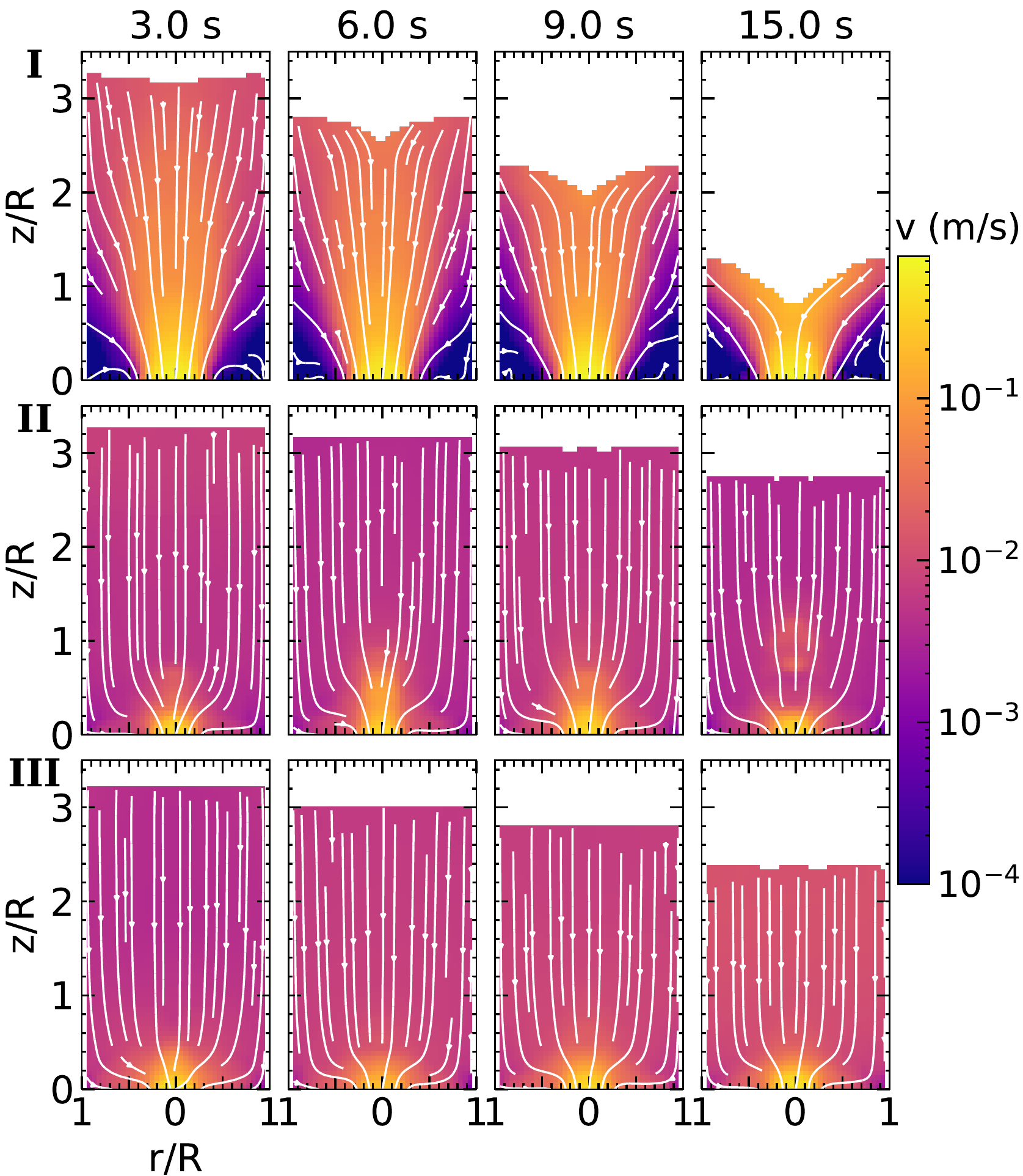}
    \caption{Coarse-grained velocity fields (excluding tangential component) inside the silo for spherocylinders with $L/d = 3.3$, next to an orifice with size $D = 54 \,\text{mm}$. The rows correspond to $f$ = 0.0, 0.32, and 1.28 Hz in order. The different columns show the velocity maps in different instants of time, as indicated above. Additionally, the streamlines have been added with white on top of the color maps.}
    \label{fig:cg_vel}
\end{figure}

The flow pattern features are also highlighted by the color maps of the magnitude of the in-plane speed $v = \sqrt{v_r^2 + v_z^2}$ (see Fig.~\ref{fig:cg_vel}). For clarity, the panels also include the streamlines of  $\boldsymbol{v}(\boldsymbol{r},t)$, drawn with the radial and the vertical components. Columns and rows represent the same cases illustrated in Fig.~\ref{fig:cg_phi}.
In the absence of external shear (stationary bottom wall), the velocity fields are rather heterogeneous, denoting a \textit{funnel flow} pattern with strong velocity gradients in the radial direction. The streamlines are considerably curved, drawing a complex flow pattern. Moreover, at the center of the silo, the material speed $|\boldsymbol{v}(\boldsymbol{r},t)|$, is significantly larger in comparison with the region close to the wall (stagnant zone), and $|\boldsymbol{v}(\boldsymbol{r},t)|$ significantly increases in the region of the orifice.

\begin{figure}
    \centering
    \includegraphics[width=\columnwidth]{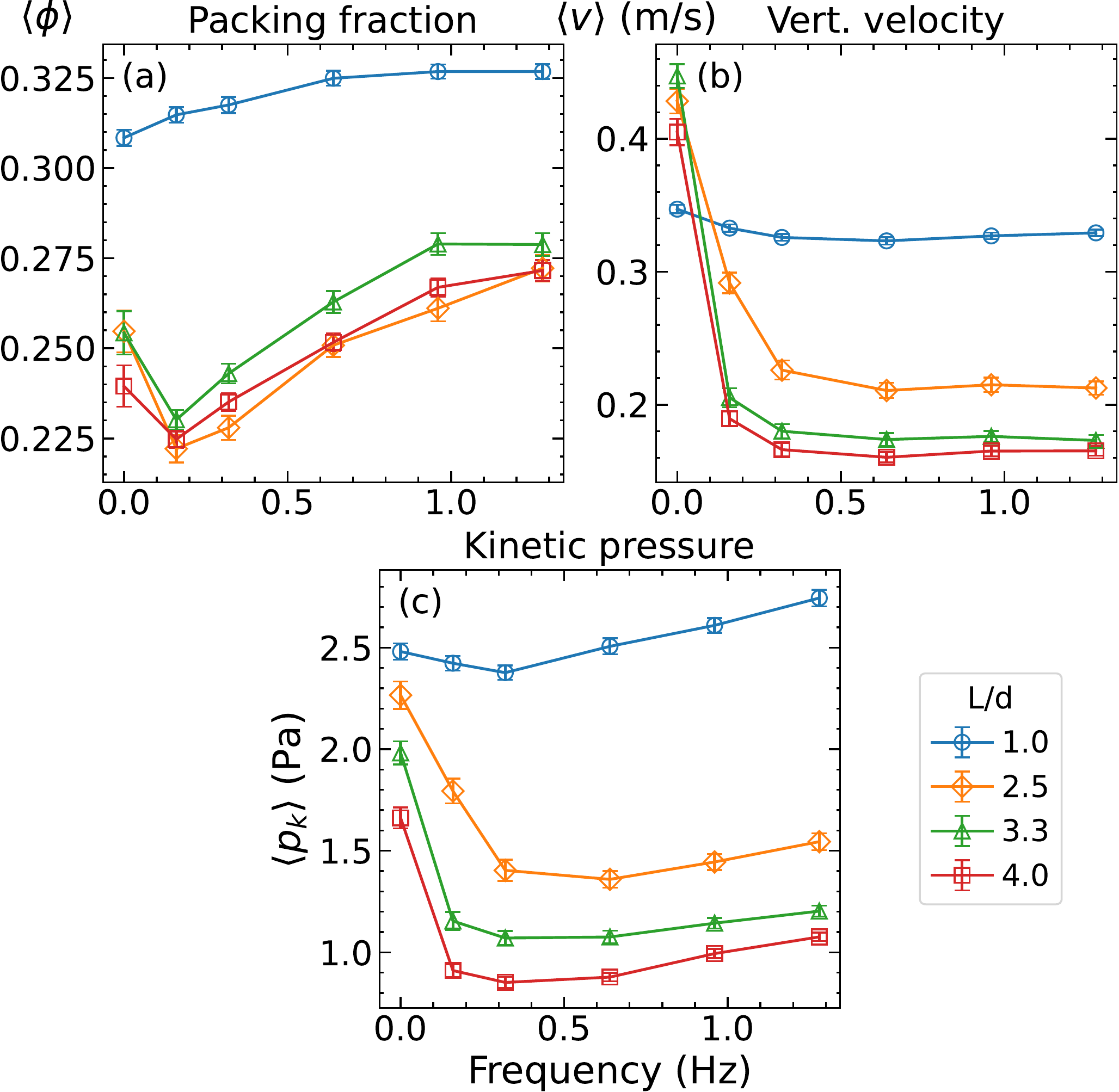}
    \caption{Macroscopic quantities averaged in the region of the orifice as a function of the rotation frequency: (a) packing fraction, (b) vertical velocity and (c) kinetic stress. Different curves correspond to different particle aspect ratios, as shown in the legend, all for the case of a fixed orifice diameter $D = 60$ mm.}
    \label{fig:phi_vel_kinstress_at_orifice}
\end{figure}

Interestingly, the appearance of the {\it mass flow} pattern seems counterintuitive, since the rotation of the bottom induces the movement of the grains in the tangential direction. It would pair with an outwards pointing fictitious force in the rotating frame. Thus, it would imply that particles go to the lateral wall of the silo, similar to what happens to water when it is put into a rotating cylinder. As our results indicate (see Fig.~\ref{fig:cg_phi}), this is not the case, and it is explained by the low magnitude of the fictitious force. Instead, we argue that the {\it mass flow} pattern emerges since the stability of the stagnant zone is affected by the rotational shear. As a result, particles are mobilized by the momentum transfer of the bottom wall, inducing the system's dilatancy and the reduction of the effective friction. Consequently, the gravity action drives the homogeneous advective particle flow, and the material moves down the silo as a column (see Fig.~\ref{fig:cg_vel}). 

Next, we thoroughly investigate the cause of the sharp decrease in $Q$, when applying the external shear, focusing on the details of the particle flow in the region of the orifice. The volumetric flow rate $Q$ through a surface is defined as the integral of the density times the velocity perpendicular to the surface $Q = \int \rho_p \phi(\boldsymbol{r}) \boldsymbol{v} (\boldsymbol{r}) \, d\boldsymbol{A} $.
Thus, to better quantify and differentiate the roles played by $\boldsymbol{v} (\boldsymbol{r})$  and $\phi(\boldsymbol{r})$ in relation to the flow rate, we have averaged these quantities in the area of the orifice during the stationary part of the flow, namely, computing the mean of the quantity $X(=v_z,\phi, O_{zz}, ...)$ using the formula $\langle X \rangle = \frac{\int X (\boldsymbol{r}) \phi (\boldsymbol{r}) \, dV}{ \int \phi (\boldsymbol{r}) \, dV} $.
Here we use $\phi (\boldsymbol{r})$ as a weight to account for the different number of particles in different cells.
The region of averaging is a 1 cm high cylinder which has its base at the orifice and its diameter is the same as the orifice size $D$. 
After this spatial integration, time and ensemble averaging is also applied.
The panels in Fig.~\ref{fig:phi_vel_kinstress_at_orifice} illustrate the dependence of $\langle \phi \rangle$ and $\langle v_z \rangle$ on the rotational frequency $f$, obtained for particles with several aspect ratios at an orifice diameter of $D = 60$ mm.
Interestingly, the evolution of $\langle \phi \rangle$ suggests a slightly increasing trend for spheres but a nonmonotonic behavior for rods, when varying $f$ (see first panel of Fig.~\ref{fig:phi_vel_kinstress_at_orifice}).
However, the magnitude of the $\langle \phi \rangle$ variation does not justify the abrupt decay of the particle flow rate of the rods. Contrarily, when raising the rotational frequency $f$ the particles notably slow down and a significant reduction of $\langle v_z \rangle$ is observed (see second panel of Fig.~\ref{fig:phi_vel_kinstress_at_orifice}). 
In fact, it strongly correlates with the trend of $Q(f)$ (see Fig.~\ref{fig:flowrate_vs_freq}), meaning that it is the larger contributing factor to the flow rate. Nevertheless the increase of the flow rate for $f > 0.3$ Hz can only be explained by the increase of $\langle \phi \rangle$.
Additionally, the third panel of Fig.~\ref{fig:phi_vel_kinstress_at_orifice} displays the evolution of the averaged kinetic pressure $\langle p_k \rangle = Tr \left( \langle \sigma^k \rangle \right)$. The latter quantity is often interpreted as a granular temperature since it is proportional to the square of the velocity fluctuations. As observed in the figure, 
the behavior of $\langle p_k \rangle$ is also in line with the obtained trend for $Q$ vs $f$, even resembling the slight nonmonotonic response reported earlier for spheres \cite{hernandez2020particle}.

\begin{figure}[t]
    \centering
    \includegraphics[width=\columnwidth]{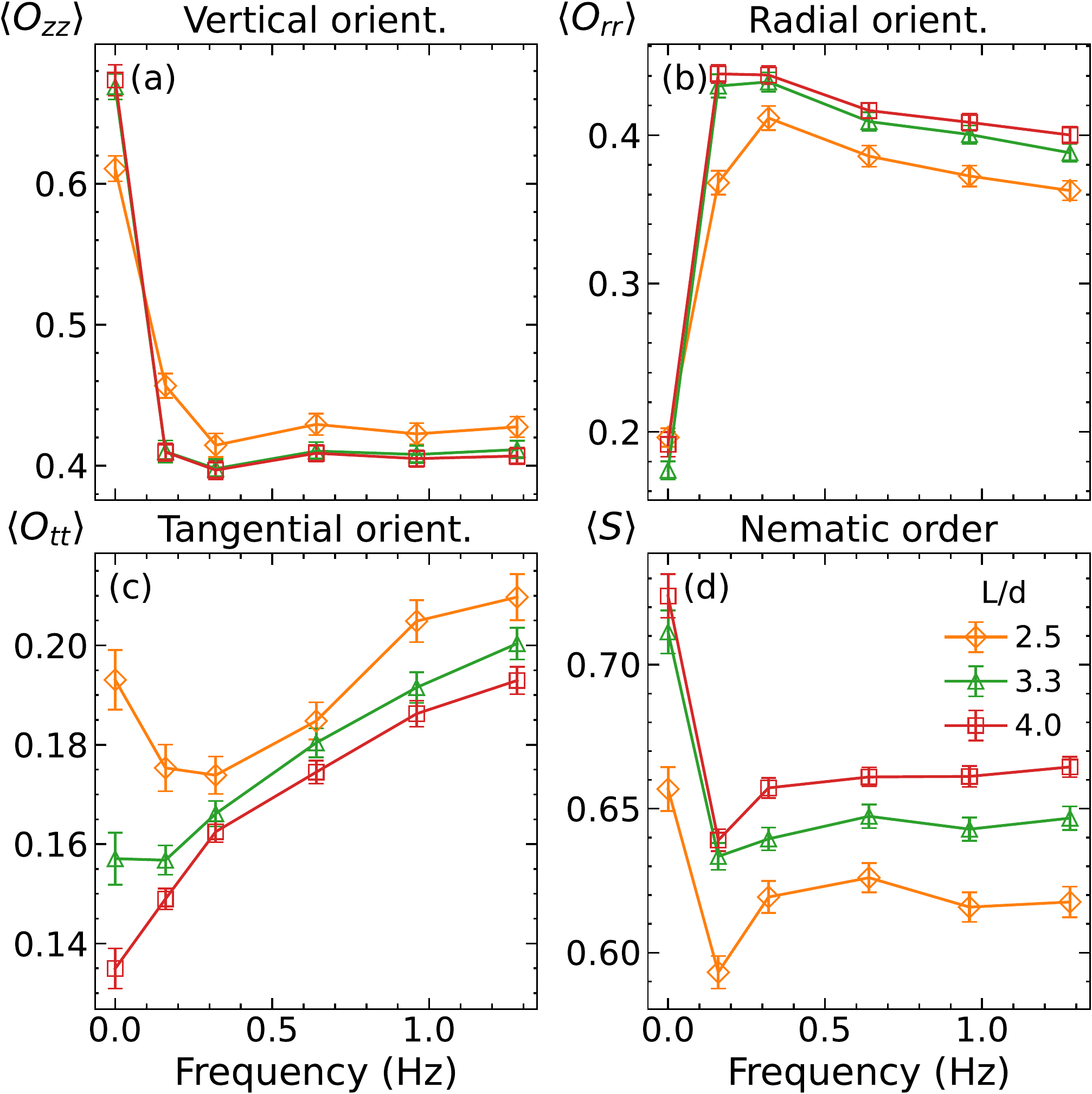}
     \caption{Mean value of the orientation quantities averaged in the region of the orifice: (a) vertical, (b) radial, (c) tangential orientation, and (d) nematic order.
     Different curves correspond to different particle aspect ratios as shown in the legend, all for the case of a fixed orifice diameter $D = 60$ mm. The inset shows the radial component of the orientation tensor.}
    \label{fig:orientations_at_orifice}
\end{figure}

Furthermore, we also inspect the influence of the rotational shear on the particle  orientation. Performing the analysis, the attention focuses on the response of the  diagonal components (in cylindrical coordinates) of the orientation tensor $\langle O_{ii} \rangle$, averaged in the region of the orifice. 
Figure \ref{fig:orientations_at_orifice} displays the diagonal elements, averaged in the same way as the quantities in Fig.~\ref{fig:phi_vel_kinstress_at_orifice}, as well as the mean nematic order $\langle S \rangle$. To obtain a meaningful measure of the actual particle order, first we averaged the orientation matrix in the region of the orifice and using the largest eigenvalue of this matrix we carried out ensemble and time averaging to acquire $\langle S \rangle$.
Interestingly, the graph indicates that in the absence of external shear ($f=0$ Hz), the rods achieve an orientation parallel to the flow direction, which is indicated by the high value of $\langle O_{zz} \rangle$, in comparison with $\langle O_{rr} \rangle$ and $\langle O_{tt} \rangle$. However, there is an abrupt change in the orientation of the particles even when a weak rotation is applied. 
Specifically, the grains seem to prefer a more horizontal (radial) direction compared to the vertically aligned case with the stationary bottom. 
Interestingly, this effect is stronger for longer particles, denoting that more elongated particles seem to face the orifice with a larger effective size.
Note that the decrease of $\langle O_{zz} \rangle$ is in the same order of magnitude as the increase in $\langle O_{rr} \rangle$ for small $f$. There are two contributing factors that can explain the latter. On one hand horizontal particles lying on the silo bottom are being pushed towards the orifice due to the mass flow pattern (see Fig.~\ref{fig:cg_vel}). On the other hand particles coming from above the orifice are more horizontally aligned \cite{supplementarymat} since they are not sheared (mass flow) until very near the orifice, while in the stationary bottom case, grains align more vertically due to the shear \cite{wegner2012alignment}.
Once again, we can argue that this tendency totally correlates with the trend of the particle flow rate (see Fig.~\ref{fig:flowrate_vs_freq}).
The orientation in the azimuthal angular direction $\langle O_{tt} \rangle$ is of low magnitude but shows a nonmonotonic or increasing trend with increasing $f$, which can be accounted to the better alignment of grains due to the external shear.
Curiously, the order of the particles $\langle S \rangle$ reduces when a slow rotation is applied but increases by further increasing the frequency, which is in correlation with the packing fraction $\langle \phi \rangle$.
These findings suggest that the level of alignment and ordering of the particles determines the macroscopic volume fraction at the orifice, impacting indirectly the macroscopic flow rate $Q$.

\begin{figure}
    \centering
    \includegraphics[width=0.9\columnwidth]{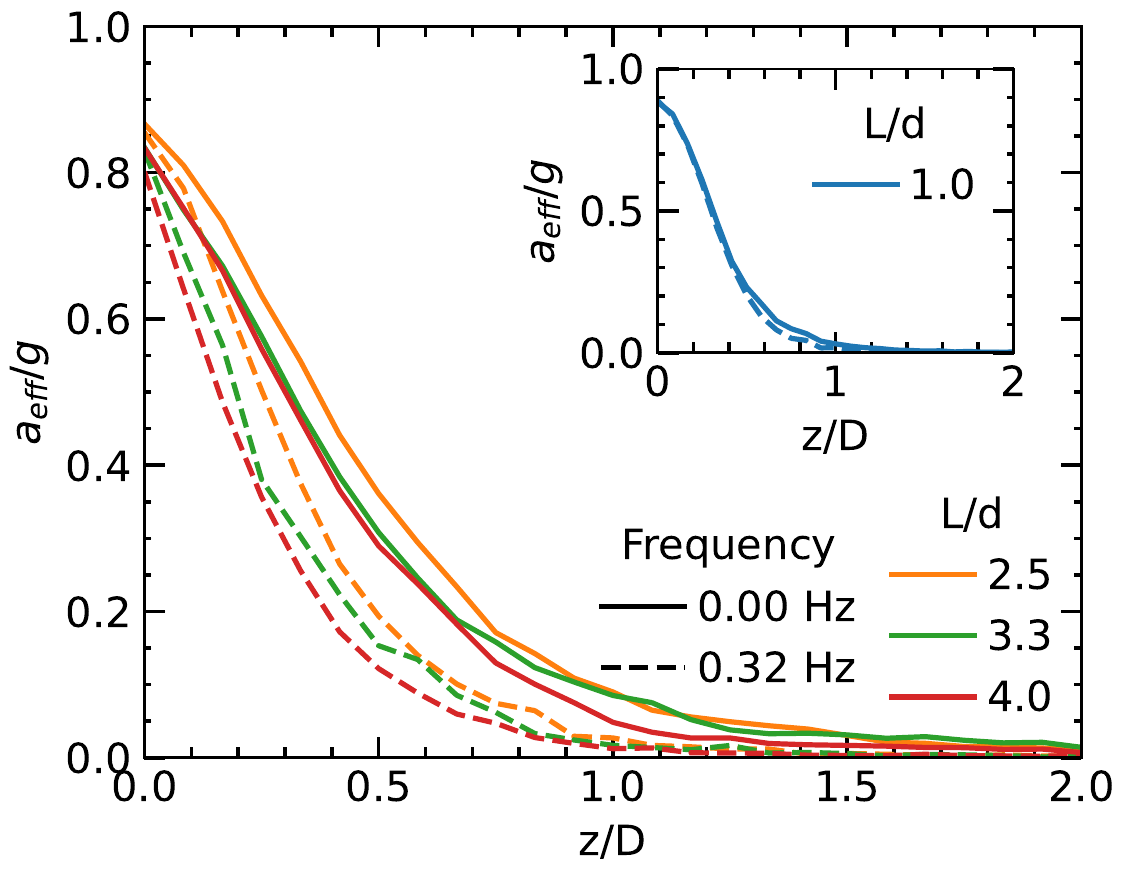}
     \caption{Relative vertical acceleration $a_{\text{eff}}/g$ of particles in the middle of the silo along the vertical axis averaged in a cylinder with diameter of 3 cm. The $x$ axis is normalized by the orifice size, which was $D = 60$ mm in this case. Solid lines correspond to stationary bottom, while the dashed lines for rotational with $f = 0.32$ Hz. The inset includes the case of the spheres for comparison.}
    \label{fig:effective_acc}
\end{figure}

In silo discharge, it is known that the particle dynamics in a distinct region above the orifice determines the exit velocity $v_z|_{_{z=0}}$ \cite{saraprl2015}.
Thus, the magnitude of $v_z|_{_{z=0}}$ can be derived from the integral of the momentum balance equation.
Figure \ref{fig:effective_acc} shows the profiles of mean acceleration $a_{\text{eff}} (z)$, obtained for several particle elongations; for comparison, the inset illustrates the data for spheres.
We define $a_{\text{eff}} (z) = \langle\frac{ \Delta v_i }{\Delta t_p} \rangle$, where  $\langle \Delta v_i \rangle$ is the mean of velocity difference of individual particles in two consecutively printed time steps ($\Delta t_p = 50$ ms). The spatial averaging was computed in the middle of the silo along the vertical axis within a cylinder with diameter $D/2$, and the time averaging during the stationary part of the flow.
The graphs include the data corresponding to a rotational velocity of $f=0.32$ Hz, and the stationary case. From these results, the message is conclusive. For flows of nonspherical particles, the rotational shear induces a sharper increase of the acceleration $a_{\text{eff}} (z)$ when approaching the exit (see dashed lines). This means that the region where the grains are accelerating is less extended for a silo under external rotational shear compared to the case of a silo with nonmoving bottom. In other words, the height of the so-called {\it free fall arch} region appears to be shrinking due to the applied rotational shear.
As a result, the areas under the curves are significantly lower than those of the static cases (see continuous lines). Remarkably, it explains the abrupt decrease of the particle velocity (and flow rate) produced by the external rotational shear, obtained for elongated grains numerically and experimentally \cite{to2021discharge}. Examining flows of spheres (see inset of Fig.~\ref{fig:effective_acc}), however, the difference between the perturbed and nonperturbed case results is notably smaller. Consequently, the mean velocity at the orifice and the particle flow rate are only affected slightly. It is worth mentioning that the same behavior is observed when exploring other rotation frequencies (see Supplemental Material \cite{supplementarymat}).

\section{Summary}
We executed a numerical analysis of the discharge of elongated particles from a silo with a rotating bottom. Interestingly, we obtain that introducing a small transverse shear might reduce the flow rate $Q$ by up to 70\% compared to a stationary bottom and by further increasing $f$, the value of $Q$ increases. We find that the relative increase is much larger in the case of rods, specifically it is larger for longer rods and smaller orifices. 
These macroscopic observations are in very good agreement with our earlier experimental findings \cite{to2021discharge}. Stepping forward, we extend the analysis, enlarging the domain of particle elongations and the effective particle diameters. Focusing attention on the dependency of the flow rate $Q$ on the orifice diameter $D$, the spheres and rods show two distinct trends. In the cases of rods, in the limit of small apertures, 
we obtain power-law relation but with a larger exponent, although in the limit of large orifices, the power-law correlation $Q \sim D^{5/2}$ is also reproduced.   
However, the obtained results are also compatible with an
exponential trend $Q\sim e^{\kappa D}$, when decreasing the orifice diameter. Very similar results have been recently obtained experimentally (see Ref.~\cite{to2021discharge}). However, it is worth mentioning that the range of examined orifice sizes is narrower than in the experiments  \cite{to2021discharge}, due to the higher probability of clogging obtained numerically. The reason behind the exponential trend obtained experimentally \cite{to2021discharge} and numerically is nontrivial. However, in the small orifice limit, the particles flow out of the silo practically one by one, and the particle flow could be understood as a stochastic process with a constant probability of passing.
Employing a coarse-graining methodology, we compute the most relevant macroscopic fields of the particle flow. Their analysis allows us to detect a transition from a funnel to a mass flow pattern as a result of the applied external shear. The averaged fields in the orifice region reveal that the initial decrease in $Q$ is mainly attributed to the velocity changes. 
Besides, the increase of the packing fraction near the orifice explains the increase in $Q$ for large rotation frequencies.
Furthermore, the flow rate correlates with the particle orientation in the vertical direction and the packing fraction with the order of grains in the region of the orifice. Finally, we also found that the vertical profiles of mean acceleration at the center of the silo are notably affected by the rotation of the bottom wall. In flows of elongated rods, it is observed that the region where the acceleration is not negligible (so-called {\it free fall arch region}) shrinks significantly due to the perturbation induced by the moving wall. 

\begin{acknowledgments}
We wish to acknowledge discussions with K.~To and E.~Somfai and financial support by the European Union's Horizon 2020 Marie Sk\l{}odowska-Curie grant ''CALIPER'' (No. 812638). R.C.~Hidalgo acknowledges the Ministerio de Ciencia e Innovación (Spanish Government) Grant PID2020-114839GB-I00 funded by MCIN/AEI/10.13039/501100011033.\smallskip

\noindent\includegraphics[width=0.09\textwidth]{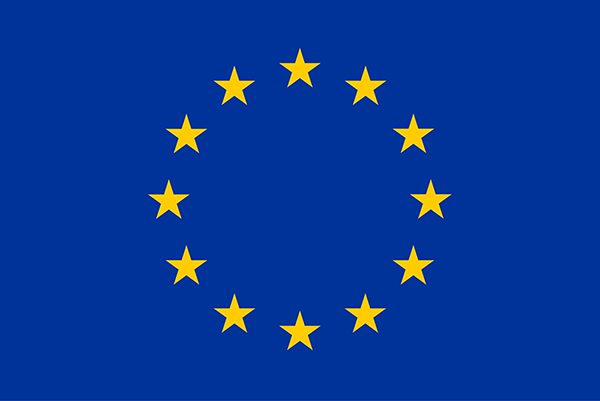}
\end{acknowledgments}

\onecolumngrid
\clearpage
\section*{Supplementary material}

\begin{figure}[htbp!]
    \centering
    (a)\includegraphics[width=0.45\textwidth]{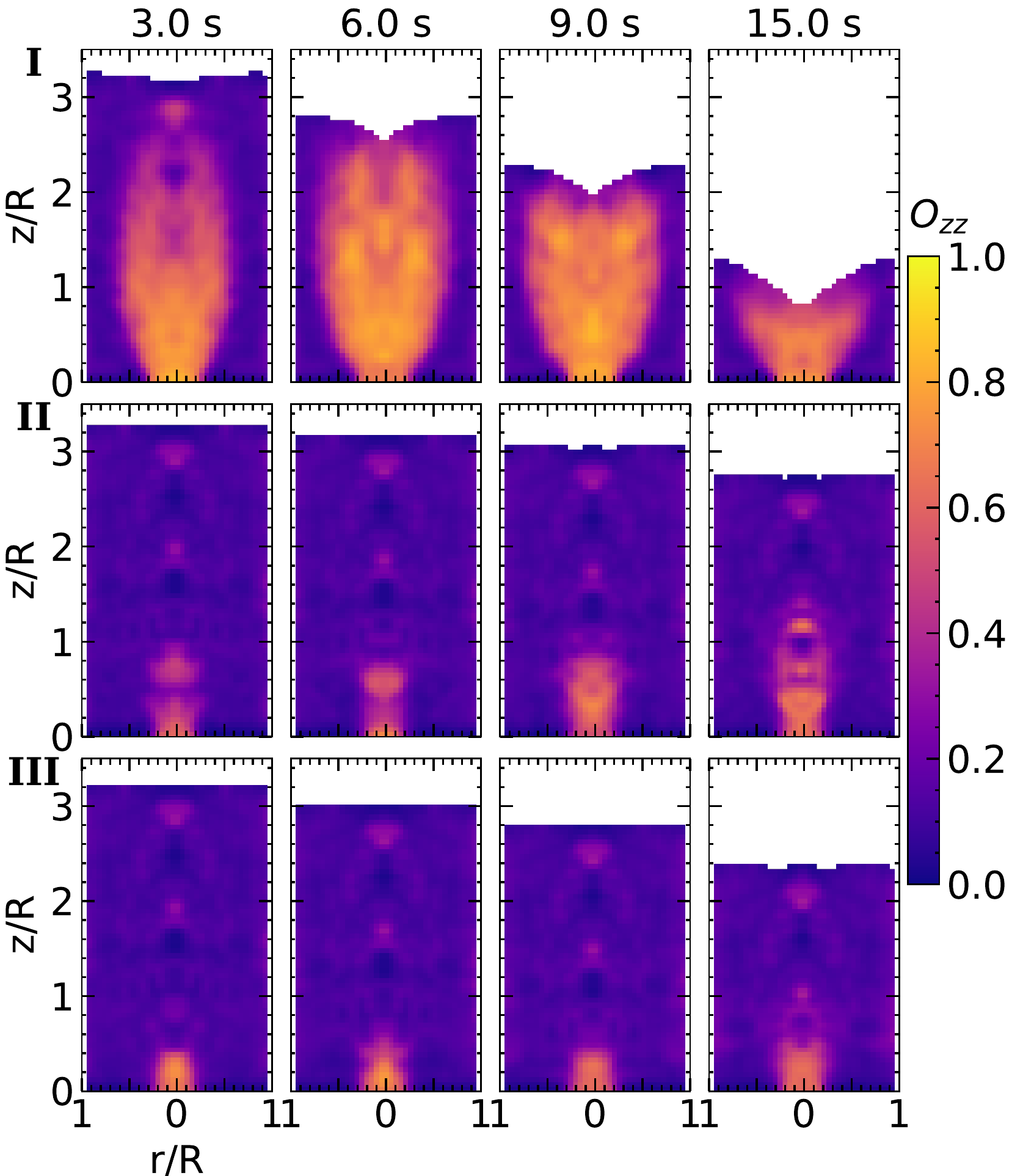}
    (b)\includegraphics[width=0.45\textwidth]{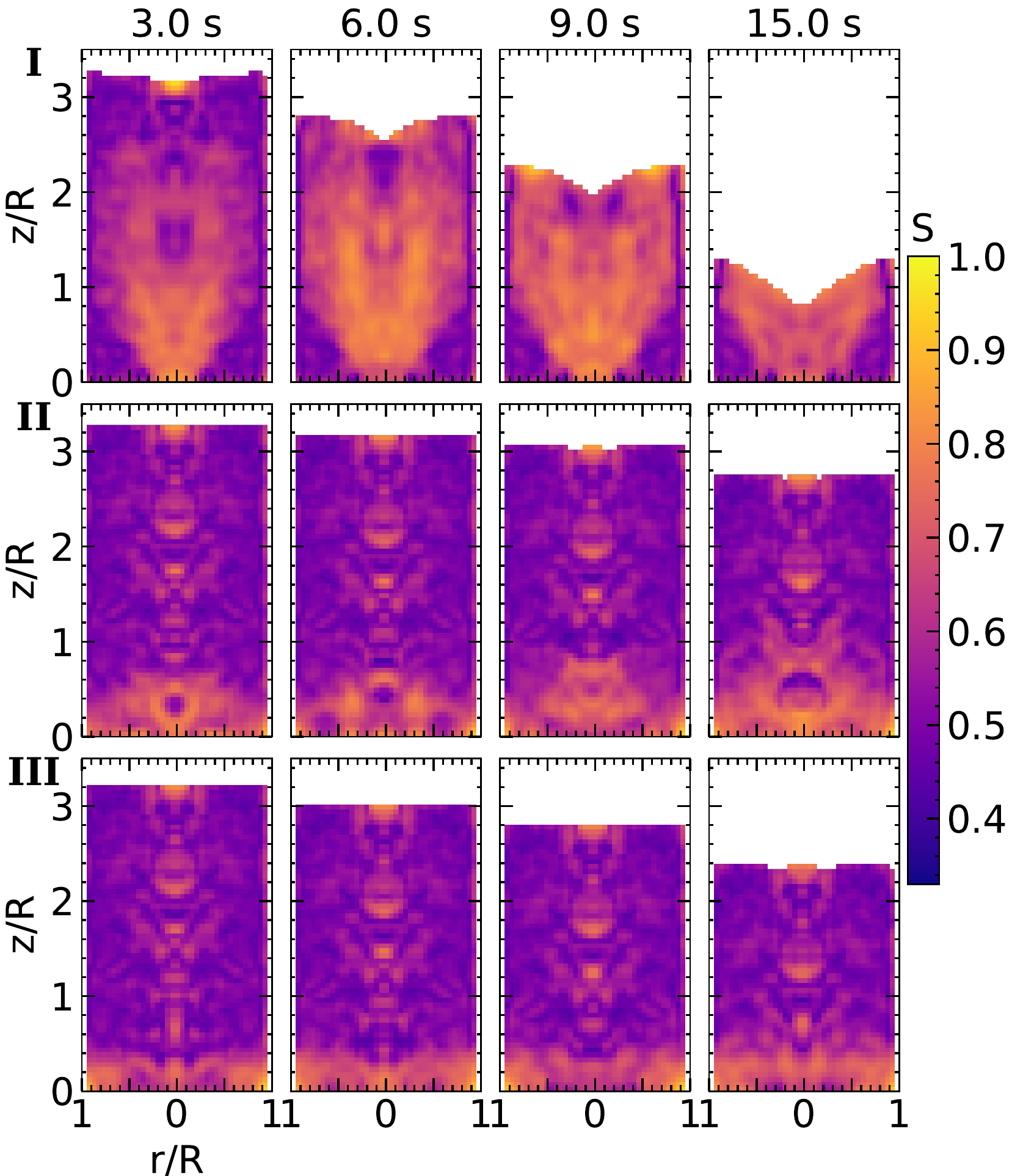}
    \caption{Color map representation of the particle vertical orientation $O_{zz}$ (a) and nematic order $S$ (b) for elongated particles of $L/d = 3.3$, next to an orifice with size $D = 54$ mm. The rows correspond to $f = 0.0, 0.32,$ and $1.28$ Hz in order. The different columns show the data in different instants of time as indicated above.}
    \label{fig:Ozz_S}
\end{figure}

\begin{figure}
    \centering
    (a)\includegraphics[width=0.45\columnwidth]{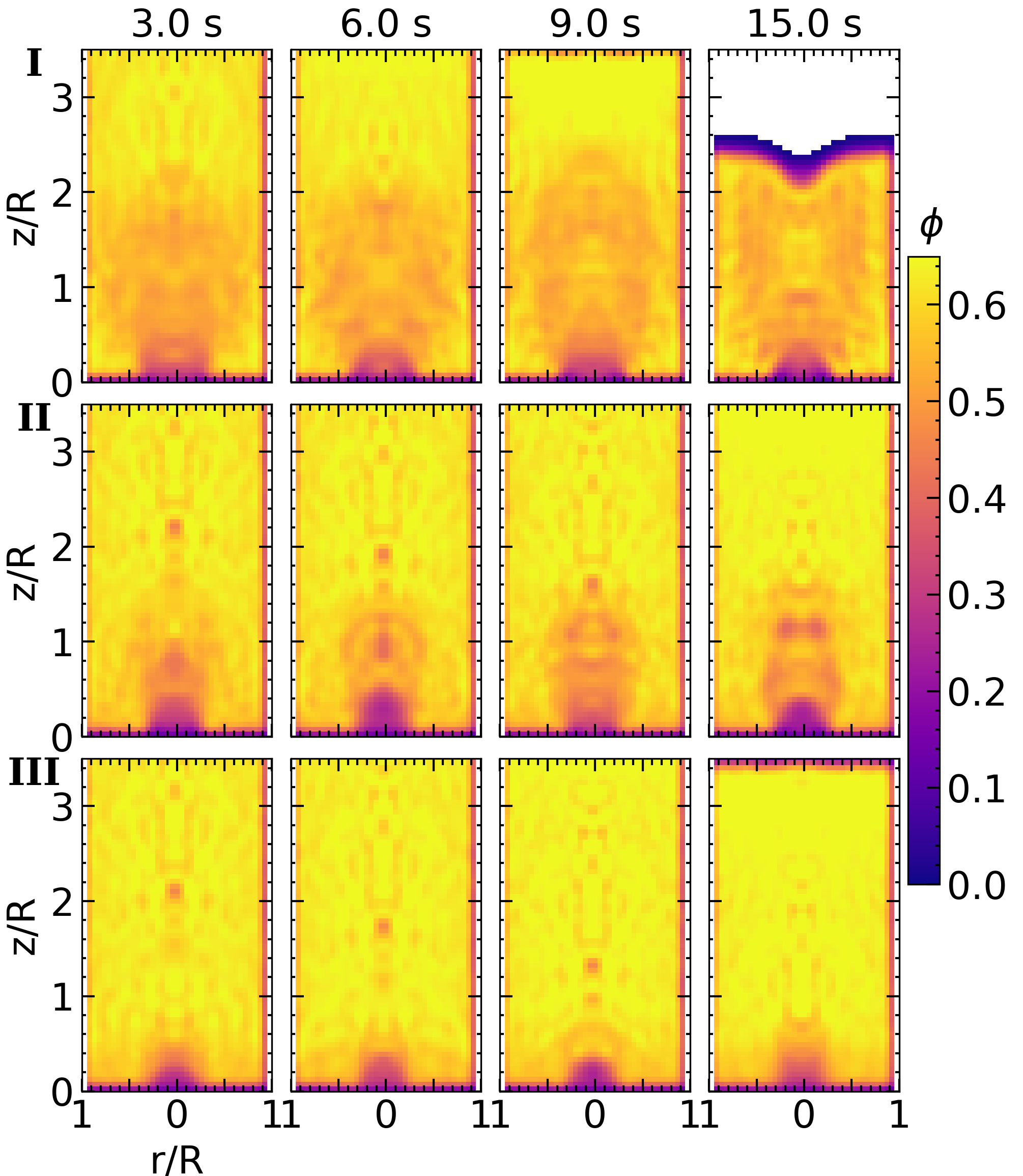}
    (b)\includegraphics[width=0.45\columnwidth]{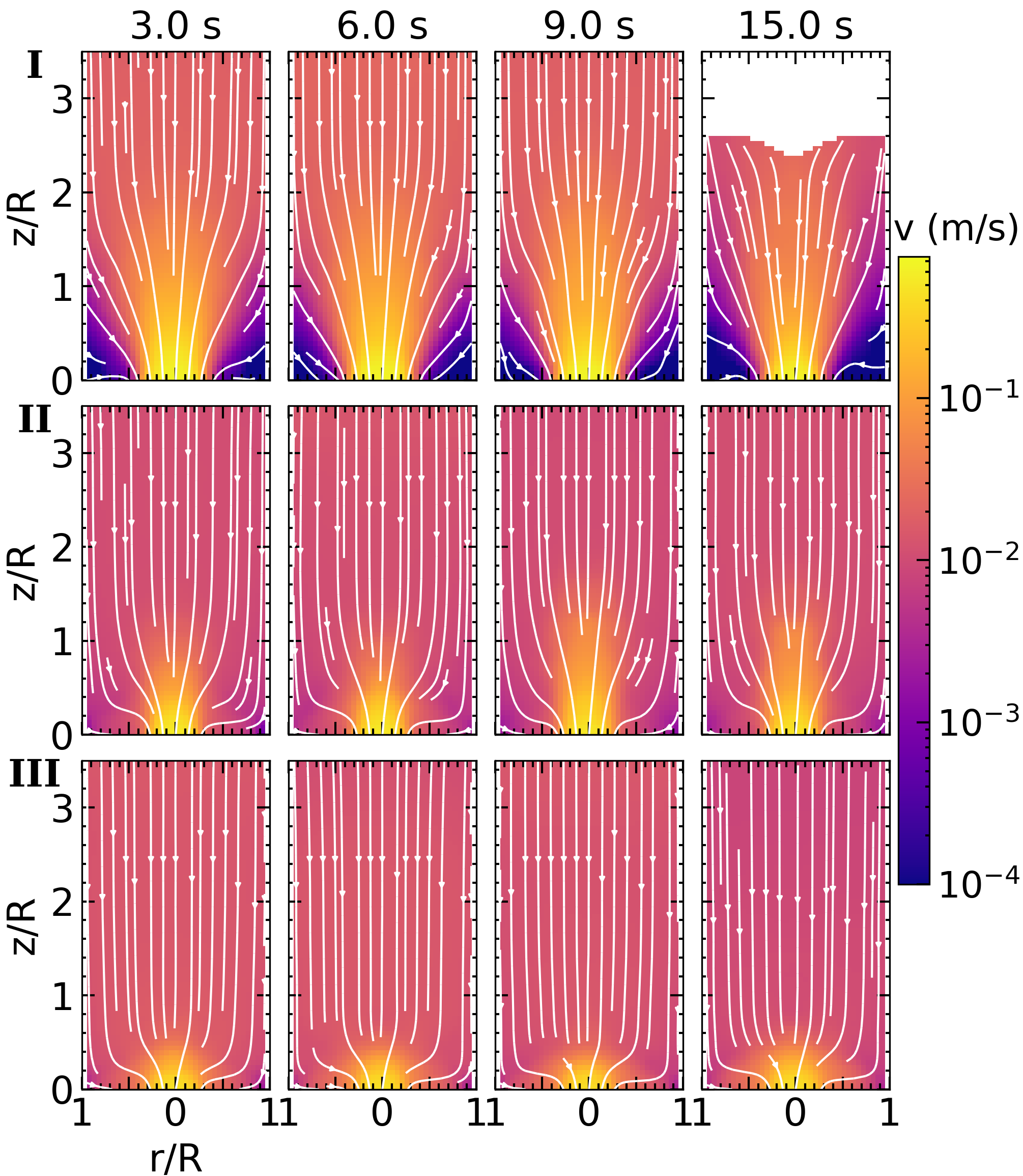}
    \caption{Color map representation of the packing fraction (a) and in-plane velocity (b) for elongated particles of $L/d = 2.5$, next to an orifice with size $D = 60$ mm. The rows correspond to $f = 0.0, 0.32,$ and $1.28$ Hz in order. The different columns show the data in different instants of time as indicated above.}
    \label{fig:Q2_5}
\end{figure}

\begin{figure}
    \centering
    (a)\includegraphics[width=0.45\columnwidth]{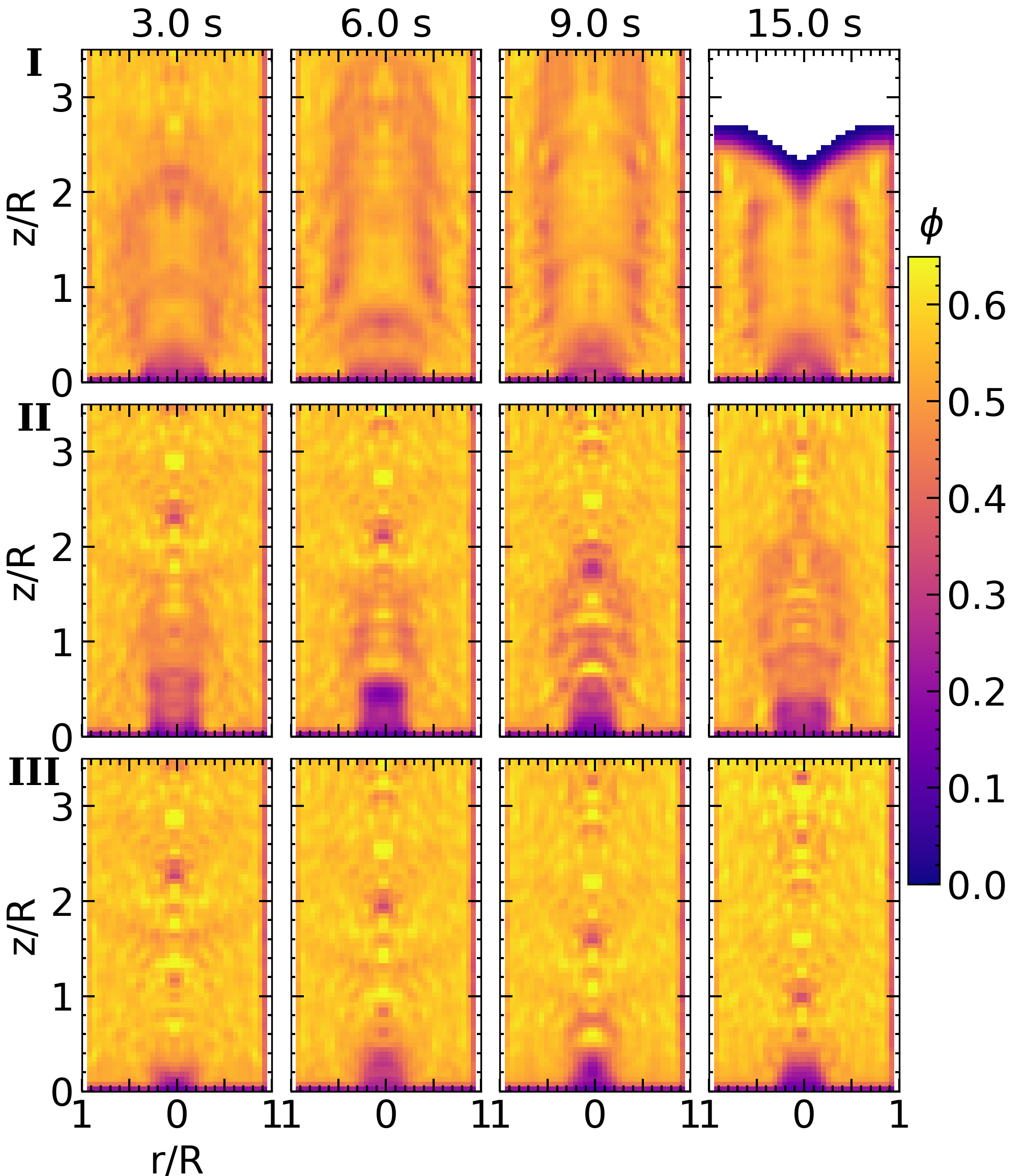}
    (b)\includegraphics[width=0.45\columnwidth]{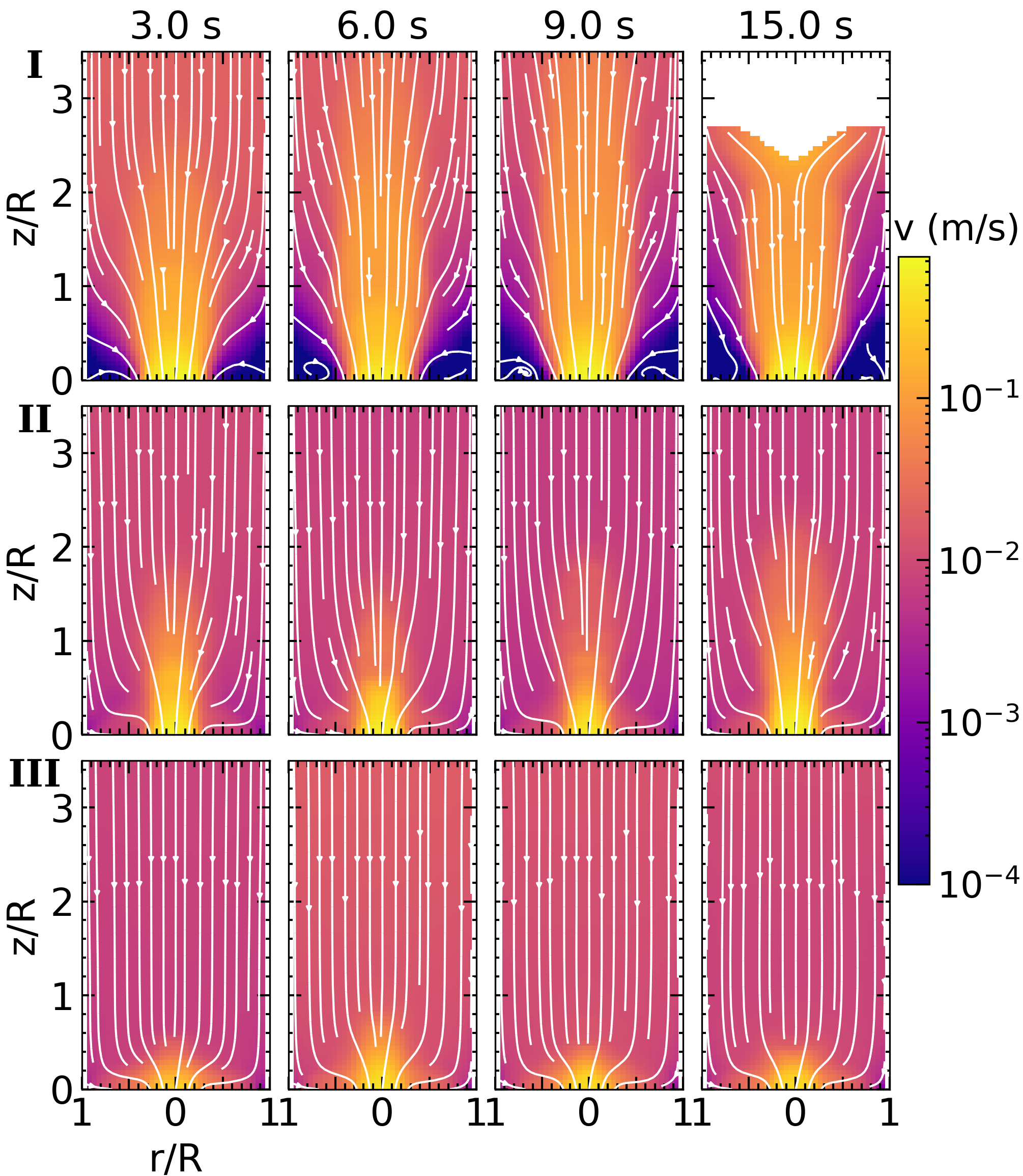}
    \caption{Color map representation of the packing fraction (a) and in-plane velocity (b) for elongated particles of $L/d = 4.0$, next to an orifice with size $D = 60$ mm. The rows correspond to $f = 0.0, 0.32,$ and $1.28$ Hz in order. The different columns show the data in different instants of time as indicated above.}
    \label{fig:Q4_0}
\end{figure}

This supplementary material includes figures that are not essential for understanding our main arguments but may provide additional insight for the interested.
The coarse-grained fields were computed as it is described in section II of the main document. 
Fig.~\ref{fig:Ozz_S}a and Fig.~\ref{fig:Ozz_S}b show the orientational field $O_{zz}$ and nematic order field $S$, respectively. The illustrated data correspond to the same cases introduced in Figure 6 of the main document.
Complementary, Figures \ref{fig:Q2_5} and \ref{fig:Q4_0} display the packing fraction and velocity fields of the discharge process for two elongations but for a slightly different orifice size ($D = 60$ mm) than what is presented in the article due to clogging in one case.


\begin{figure}
    \centering
    \includegraphics[width=0.45\columnwidth]{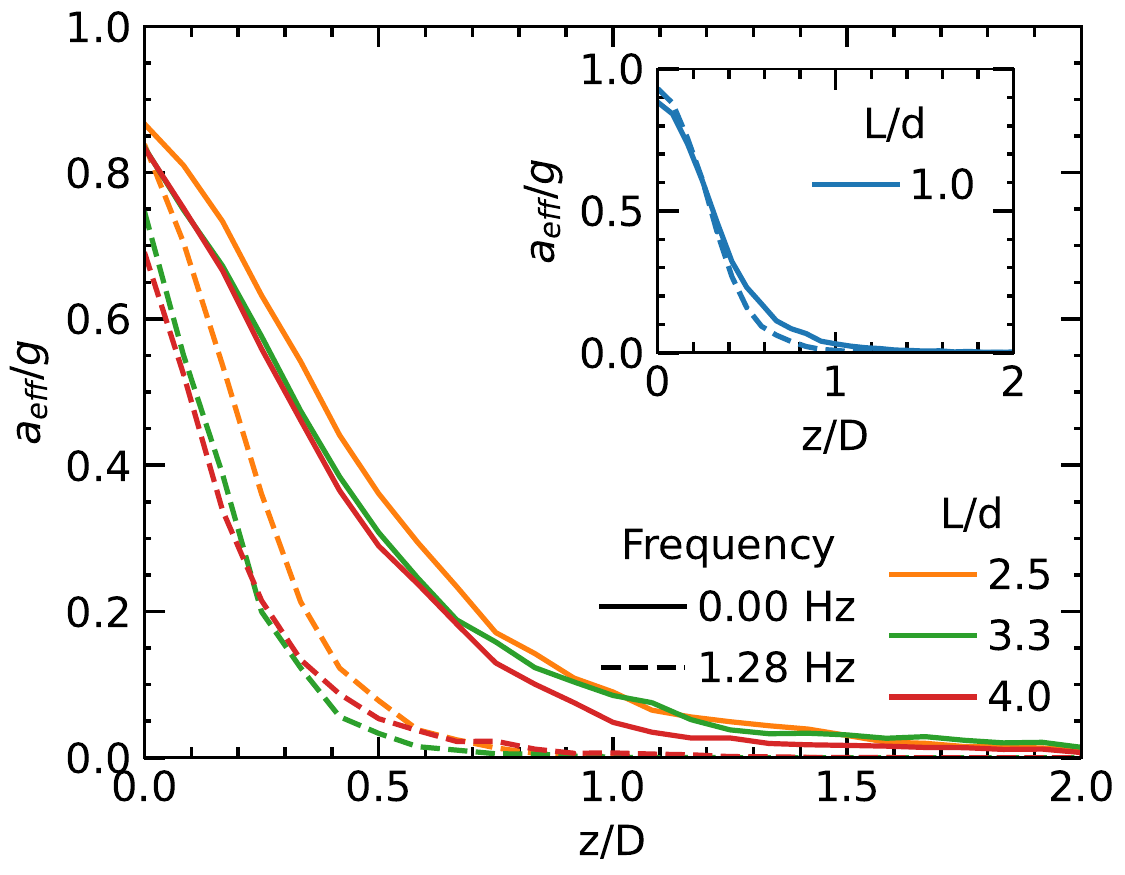}
    \caption{Relative vertical acceleration $a_{\rm eff}/g$ of particles in the middle of the silo along the vertical axis averaged in a cylinder with diameter of 3 cm. The x-axis is normalized by the orifice size, which was $D = 60$ mm in this case. Solid lines correspond to stationary bottom, while the dashed lines for rotational with $f = 1.28$ Hz. The inset includes the case of the spheres for comparison.}
    \label{fig:a_eff}
\end{figure}

\end{document}